\documentclass[conference,10pt]{IEEEtran}

\usepackage{cite}

\ifCLASSINFOpdf
  \usepackage[pdftex]{graphicx}
\else
\fi

\usepackage{amsmath}

\usepackage{multirow}
\usepackage{enumitem}
\usepackage{listings}
\newcommand{\TT}[1]{\texttt{#1}}
\newcommand{\BF}[1]{\textbf{#1}}

\begin{document}

\title{Verifying High-Level Latency-Insensitive Designs with Formal Model Checking\vspace{-8pt}}

\author{
\IEEEauthorblockN{Steve Dai,
Alicia Klinefelter,
Haoxing Ren, 
Rangharajan Venkatesan,\\
Ben Keller, 
Nathaniel Pinckney,
Brucek Khailany}
\IEEEauthorblockA{NVIDIA}
\vspace{-26pt}
}

\maketitle

\begin{abstract}
Latency-insensitive design mitigates increasing interconnect delay and enables productive component reuse in complex digital systems.
This design style has been adopted in high-level design flows because untimed functional blocks connected through latency-insensitive interfaces provide a natural communication abstraction.
However, latency-insensitive design with high-level languages also introduces a unique set of verification challenges that jeopardize functional correctness.
In particular, bugs due to invalid consumption of inputs and deadlocks can be difficult to detect and debug with dynamic simulation methods.
To tackle these two classes of bugs, we propose formal model checking methods to guarantee that a high-level latency-insensitive design is unaffected by invalid input data and is free of deadlock.
We develop a well-structured verification wrapper for each property to automatically construct the corresponding formal model for checking.
Our experiments demonstrate that the formal checks are effective in realistic bug scenarios from high-level designs.
\end{abstract}

\IEEEpeerreviewmaketitle

\section{Introduction}

As modern SoC design challenges continue to motivate reuse of existing design blocks, latency-insensitive (LI) design has emerged as a practical methodology for synchronizing pre-assembled modules under increasing pressure of lengthening interconnect delay~\cite{carloni2003methodology,carloni2015latency}.
By exposing a valid-ready interface from each module, LI design decouples the timing of intra-module computation from that of inter-module communication to ensure robust functionality while tolerating arbitrary communication latency between modules.
This methodology enables flexible physical design implementation without impacting verification of individual components.

In parallel with this trend, hardware designers have embraced high-level languages for high-productivity VLSI design.
In particular, high-level synthesis (HLS) compilers can automatically synthesize RTL from C++ models.
Because an HLS compiler translates untimed functional blocks in software into interconnected cycle-accurate hardware modules with customized throughput and latency, it is natural for HLS to adopt an LI-based composition of the modules to take advantage of the modularity and relaxed timing requirement of LI design.
The confluence of HLS and LI design has enabled rapid design of large-scale chips using high-level languages~\cite{venkatesan20190,ajayi2017celerity}.

\begin{figure}[t]
\centering
\includegraphics[trim=20pt 350pt 238pt 0pt, width=1.03\columnwidth]{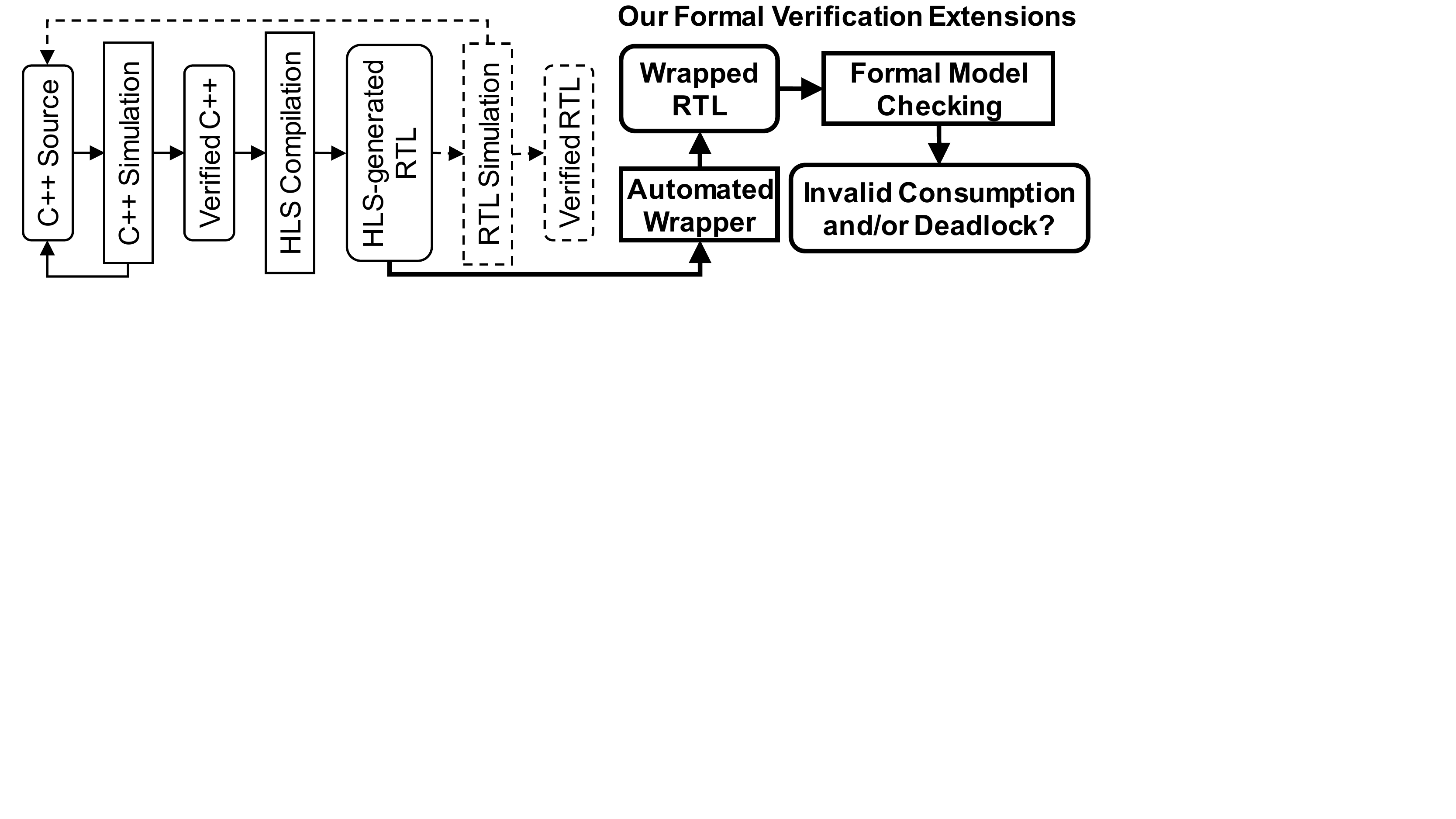}
\caption{\BF{Typical HLS design and verification flow with our proposed formal extensions} --- Design is typically verified with dynamic simulations in C++ and RTL. We extend verification with automated formal model checking.\vspace{3pt}}
\label{fig:hls_flow}
\end{figure}

Figure \ref{fig:hls_flow} shows a typical HLS design flow (on the left) for generating an RTL model from a C++ description.
Designers generally rely on dynamic simulation of C++ and RTL models for verifying their designs.
The HLS design flow allows designers to leverage C++ simulation for the bulk of their design verification tasks and promises orders of magnitude speedup over a conventional RTL flow~\cite{cong2011high}.
Nevertheless, due to inherent differences in timing models between C++ and RTL, C++ simulation must be complemented by RTL simulation to expose bugs that require cycle-accurate introspection.

The quality of dynamic simulation depends heavily on the effectiveness of the set of chosen stimuli. These stimuli  realistically represent only a small subspace of all possible inputs and risk excluding difficult-to-anticipate corner cases.
In addition, the LI property allows for arbitrary timing of inter-module interface, further requiring expansion of the verification coverage space to include a full range of input arrival times and relative input ordering.
With dynamic simulation, it is especially difficult to detect and eliminate design issues resulting in invalid consumption of inputs and deadlock, which constitute two commonly occurring bugs in LI-based HLS design.  To address these two classes of bugs, we propose augmenting a typical HLS design flow with formal verification extensions (right side of Figure \ref{fig:hls_flow} in bold).
The major contributions of this work are:
\begin{itemize}[leftmargin=*]
\item We propose RTL formal model checking methods that guarantee an LI-based HLS design is not affected by invalid input data and is free of deadlock for all intended use cases.
\item We develop an automated flow to generate the verification wrapper for each property and implement the corresponding formal checks without human intervention.
\item We demonstrate the effectiveness of the proposed verification models and flow in realistic industry and academic bug scenarios, as well as on a range of design blocks. 
\end{itemize}

The remainder of the paper is organized as follows: Section~\ref{sec-related} reviews related work;
Section~\ref{sec-li} establishes the basics for LI design in HLS; 
Section~\ref{sec-formal} describes the two classes of common bugs we target and proposes two formal models to detect these bugs; Section~\ref{sec-experiments} demonstrates the effectiveness of our models, followed by conclusions in Section~\ref{sec-conclusions}.

\section{Related Work}
\label{sec-related}

LI design refers to the correct-by-construction composition of stallable computational processes that exchange data through communication channels in accordance with an appropriate LI protocol such as valid/ready or request/acknowledge~\cite{carloni2003methodology,carloni2001theory}.
As long as each computational process (i.e., functional module) is implemented correctly, the composition will also behave correctly regardless of the latency of the channels.
The introduction of the methodology has led to the emergence of a family of LI protocols, followed by a set of dynamic simulation and formal verification techniques to validate the correctness of these protocols~\cite{suhaib2006validating,li2007design}.
Our work is concerned with verifying designs implemented using LI methodology, rather than validating the correctness of the methodology and protocols as in these previous works.

Designers typically verify LI-based HLS designs with dynamic simulation in C++ and RTL as they iterate on various design changes.
Conventionally, formal verification methods have been applied to prove the correctness of C-to-RTL transformations by checking the equivalence of the model before and after transformation, rather than validating the correctness of the C++ implementation itself~\cite{karfa2008equivalence,kim2004automated}.
KAIROS leverages formal equivalence checking to verify whether a modified HLS design is equivalent to the original (golden) design after incremental code modification or change in HLS optimizations~\cite{piccolboni2019kairos}.
Unlike previous formal techniques for HLS, our methods do not target bugs caused by the transformations of the HLS tool.
Instead, we target designer-induced bugs while not requiring a golden model as ground truth.

\section{LI Implementation in HLS}
\label{sec-li}

In this work, LI-based HLS designs are realized using MatchLib, a high-level library of synthesizable port and channel primitives in C++ implementing LI valid-ready protocols for HLS tools~\cite{khailany2018modular}.
For an LI design in HLS, each functional block exposes an interface of directional ports from the MatchLib library, shown as \TT{InA}, \TT{InB}, and \TT{Out} in Figure~\ref{code:simple_adder}(a). 
These ports are then connected to MatchLib ports on other functional blocks via MatchLib channels.
As shown in Figure~\ref{code:simple_adder}(a), each port can read (pop) data from a channel (e.g., \TT{InA} and \TT{InB}) or write (push) data to a channel (e.g., \TT{Out}).

\begin{figure}[t]
\begin{minipage}[c]{0.45\columnwidth}
  \begin{minipage}[l]{0.05\columnwidth}
  \hspace{0.05\columnwidth}      
  \end{minipage}  
  \begin{minipage}[l]{0.9\columnwidth}
    \centering
    \small
\begin{lstlisting}[language=C++,columns=fullflexible]
void adder_blocking() {
 while (1) {
  DataA = InA.Pop();
  DataB = InB.Pop();
  Out.Push(DataA+DataB);
}}
\end{lstlisting}
    \vspace{-3pt}(a)
  \end{minipage}
  \\
  \begin{minipage}[l]{0.83\columnwidth}
  \vspace{13pt}
  \caption{\BF{Simple adder design in C++} --- (a) Input ports are blocking. (b) Input ports are non-blocking.}
  \label{code:simple_adder}
  \end{minipage}
  \begin{minipage}[r]{0.01\columnwidth}
  \hspace{0.05\columnwidth}      
  \end{minipage}
\end{minipage}
\begin{minipage}[c]{0.5\columnwidth}
  \begin{minipage}[r]{0.05\columnwidth}
  \hspace{0.05\columnwidth}      
  \end{minipage} 
\begin{minipage}[r]{0.99\columnwidth} 
    \centering
    \small
\begin{lstlisting}[language=C++,columns=fullflexible]
void adder_nonblocking() {
 statusA = false; statusB = false;
 while (1) {
  if (!statusA)
   statusA = InA.PopNB(DataA);
  if (!statusB)
   statusB = InB.PopNB(DataB);
  if (statusA && statusB) {
   Out.Push(DataA+DataB); 
   statusA = false; statusB = false;
}}}
\end{lstlisting}
    \vspace{-5pt}(b)
\end{minipage}
\end{minipage}
\end{figure}

Transactions acting on a MatchLib port can be either blocking or non-blocking.
Blocking communication prevents subsequent transactions from executing until the current read or write has succeeded, and can block forward progress.
In contrast, non-blocking communication allows subsequent transactions to execute regardless of whether the current read or write is successful, and cannot block forward progress.
Figure~\ref{code:simple_adder}(b) implements the same adder design as in Figure~\ref{code:simple_adder}(a), except with non-blocking input ports \TT{InA} and \TT{InB}.
\TT{statusA} and \TT{statusB} in (b) are used to indicate whether \TT{DataA} and \TT{DataB} respectively contain valid input data from the channels.
Because of the use of a blocking pop in (a), Line 4 cannot be executed until the \TT{Pop} in Line 3 successfully reads valid data from port \TT{InA} into local variable \TT{DataA}.
However, with non-blocking pop in (b), Line 6 can be executed after Line 5 regardless of whether the \TT{PopNB} in Line 5 is successful.
The corresponding valid-ready interfaces in RTL for input port \TT{InA} and output port \TT{Out} generated from the C++ descriptions in Figure~\ref{code:simple_adder} are shown in Figure~\ref{code:simple_adder_interface}, respectively.
Blocking and non-blocking ports share the same RTL interface but differ in their internal blocking logic.

\begin{figure}[t]
  \begin{minipage}[c]{0.05\columnwidth}
  \hspace{0.05\columnwidth}      
  \end{minipage}%
  \begin{minipage}[c]{0.45\columnwidth}
    \centering
    \small
\begin{lstlisting}[language=Verilog,columns=fullflexible]
input [31:0] InA_msg;   
input InA_val;                
output InA_rdy;
\end{lstlisting}
  \end{minipage}%
  \begin{minipage}[c]{0.45\columnwidth}
    \centering
    \small
\begin{lstlisting}[language=Verilog,columns=fullflexible]
output [31:0] Out_msg;
output Out_val;
input Out_rdy;
\end{lstlisting}
  \end{minipage}%
\caption{\BF{Interface of simple adder design in Verilog after HLS} --- Valid/ready input port signals and output port signals, respectively.}
\label{code:simple_adder_interface}
\end{figure}

\begin{figure}[b]
\begin{minipage}[l]{0.4\columnwidth}
\centering
\includegraphics[trim=0.05in 20pt 200pt 0.05in, clip, width=0.8\columnwidth]{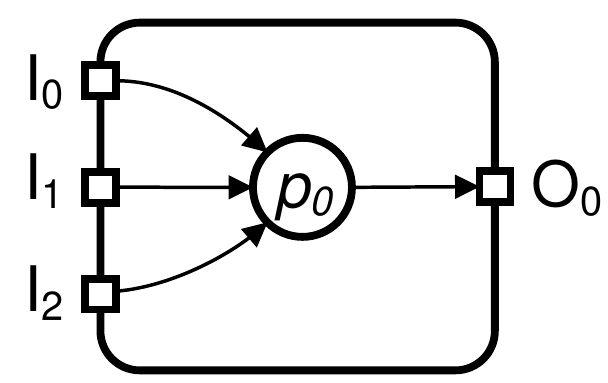}
\end{minipage}
\begin{minipage}[l]{0.57\columnwidth}
\caption{\BF{Dataflow for a three-input one-output process} --- Internal details of process $p_0$, in addition to the dataflow model, determines whether input ports should be blocking or non-blocking.}
\label{fig:process_compositions}
\end{minipage}
\end{figure}

It is important for an LI implementation to include non-blocking communication so that realistic and scalable designs can be expressed.
Figure~\ref{fig:process_compositions} illustrates a three-input one-output process $p_0$.
If $p_0$ implements a multiplier, all three inputs must be available before the output is valid; however, if $p_0$ is an arbiter, only one input needs to be available before producing a valid output. 
The multiplier could be implemented with blocking communication, while the arbiter must include non-blocking reads to be functional. 

While the decision to use blocking versus non-blocking communication is dependent on internal process details, each comes with its own pitfalls that result in the two classes of bugs targeted by this paper.
Blocking communication creates an inherent wait-for relationship between design blocks and can lead to deadlock.
Non-blocking communication requires custom bookkeeping logic (e.g., Lines 4, 6, and 8 in Figure~\ref{code:simple_adder}(b)) to prevent undesirable consumption of invalid input data that can increase the chance of designer error.
Mixing blocking and non-blocking communication further complicates the design.
Our work provides safeguards from these kinds of unintended and undesirable consequences through automation, without extra burden on the HLS designer or verification engineer.

\section{Formal Model Checking for LI Design}
\label{sec-formal}

The first class of bugs we target involves consumption of invalid inputs at the LI interface.
When non-blocking communication is needed, a user typically writes custom bookkeeping logic to manage the communication.  This is a common cause of errors, since improperly constrained non-blocking reads and writes can result in tainted updates to stateful elements by invalid input data.
The risk of error increases with additional non-blocking ports as designers attempt to manage the complex interaction among multiple instances of custom bookkeeping logic while keeping track of how the sequential C++ design entry will translate into parallel hardware in generated RTL.
This class of bugs is difficult to detect with dynamic simulation because the designer-imposed constraints are often buggy only under limited corner cases that are non-trivial to conceive ahead of time during test planning.

The second class of bugs we target involves deadlock, which may arise due to a multitude of factors, including incorrect capacities for communication channels, improper application or combination of blocking and non-blocking ports, latency-sensitive bookkeeping logic, or circular dependencies among different blocks.
It is difficult for designers to be completely mindful of all sources of deadlock during the design process, especially when the design entry is untimed and sequential C++.
Exposing potential deadlock scenarios with dynamic simulation requires stressing the range of input arrival times (and therefore relative input orderings) at the LI interfaces within the design.
Unfortunately, finding buggy combinations of signal arrivals can require many iterations to expose bugs requiring complex input patterns.

Formal model checking~\cite{clarke2018model} is commonly applied to more thoroughly verify various hardware components and protocols~\cite{bingham2011parameterized,kaufmann2019verifying}. 
Therefore, it is uniquely positioned to address the verification gap by proving the absence of the two classes of bugs in our designs without limitation to a specific subset of stimulus and input arrival timing.
In particular, we apply RTL-based formal model checking on the HLS-generated RTL design blocks to verify properties associated with these two classes of bugs.
Figures~\ref{fig:invalid_check} and~\ref{fig:deadlock_check} present an overview of the two corresponding formal models.

\subsection{Invalid Input Consumption Check}
\label{subsec-invalid}

\begin{figure}[t]
\centering
\includegraphics[trim=18pt 283pt 183pt 0pt, width=\columnwidth]{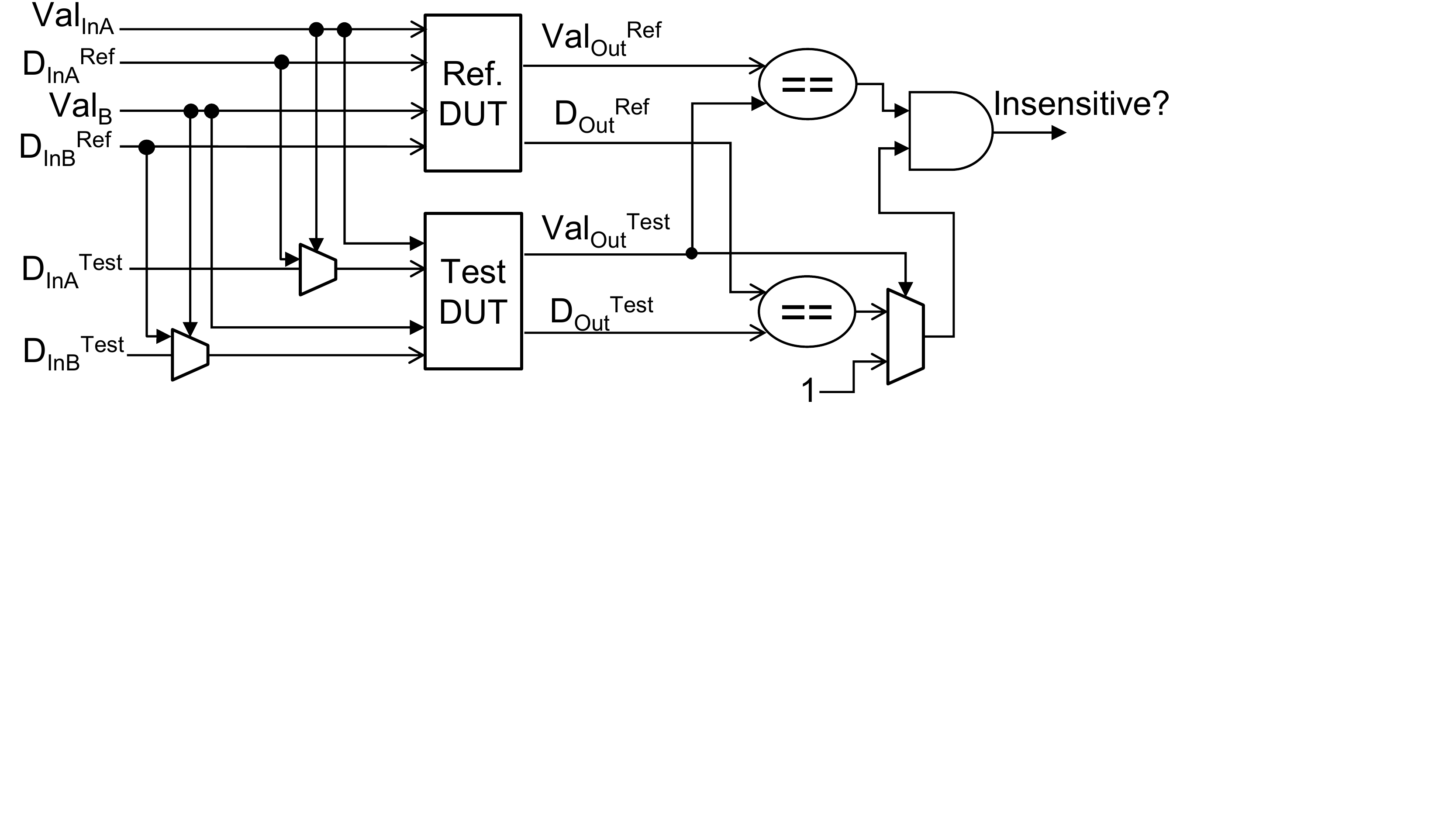}
\caption{\BF{Formal model for verifying invalid input insensitive property} --- Verifies if valid outputs are equivalent even under different invalid inputs.}
\label{fig:invalid_check}
\end{figure}

Our invalid input consumption check proves whether valid output data of the design are unaffected by invalid input data.
In other words, invalid input data must not assert influence on any valid output data.
Recall the non-blocking adder in Figure~\ref{code:simple_adder}(b) in which input data are read from non-blocking ports \TT{InA} and \TT{InB}. 
If Line 9 in Figure~\ref{code:simple_adder}(b) is not guarded by the conditional statement in Line 8, the adder will perform addition regardless of whether \TT{DataA} and \TT{DataB} contain valid input data.
In this case, valid output at \TT{Out} is affected by invalid inputs, and fails the property of being unaffected by invalid input.
While this hypothetical bug represents a relatively contrived case of incorrect bookkeeping logic, there are many examples of improperly constrained non-blocking operations that can be discovered by this check.

Figure~\ref{fig:invalid_check} shows how we wrap a design under test (DUT) into a formal model that checks for invalid input consumption.
This model consists of a reference as well as a test instance of the same DUT, with corresponding reference and test inputs and outputs as shown.
On the input side, both DUT instances are set up to always receive the same valid inputs, but may receive different invalid inputs, as modeled by the multiplexers.
On the output side, the model verifies that any valid outputs are equivalent by comparing the corresponding output valid and output data signals.
$Val^{Ref}==Val^{Test}$ and $Val^{Ref}\& Val^{Test}\implies D^{Ref}==D^{Test}$ define the conditions that constrain the inputs and check the outputs.
Note that the model is set up such that both the reference and test DUTs receive the same external ready signal.

\subsection{Deadlock Detection}
\label{subsec-deadlock}

\begin{figure}[t]
\centering
\includegraphics[trim=0pt 303pt 468pt 0pt, width=0.68\columnwidth]{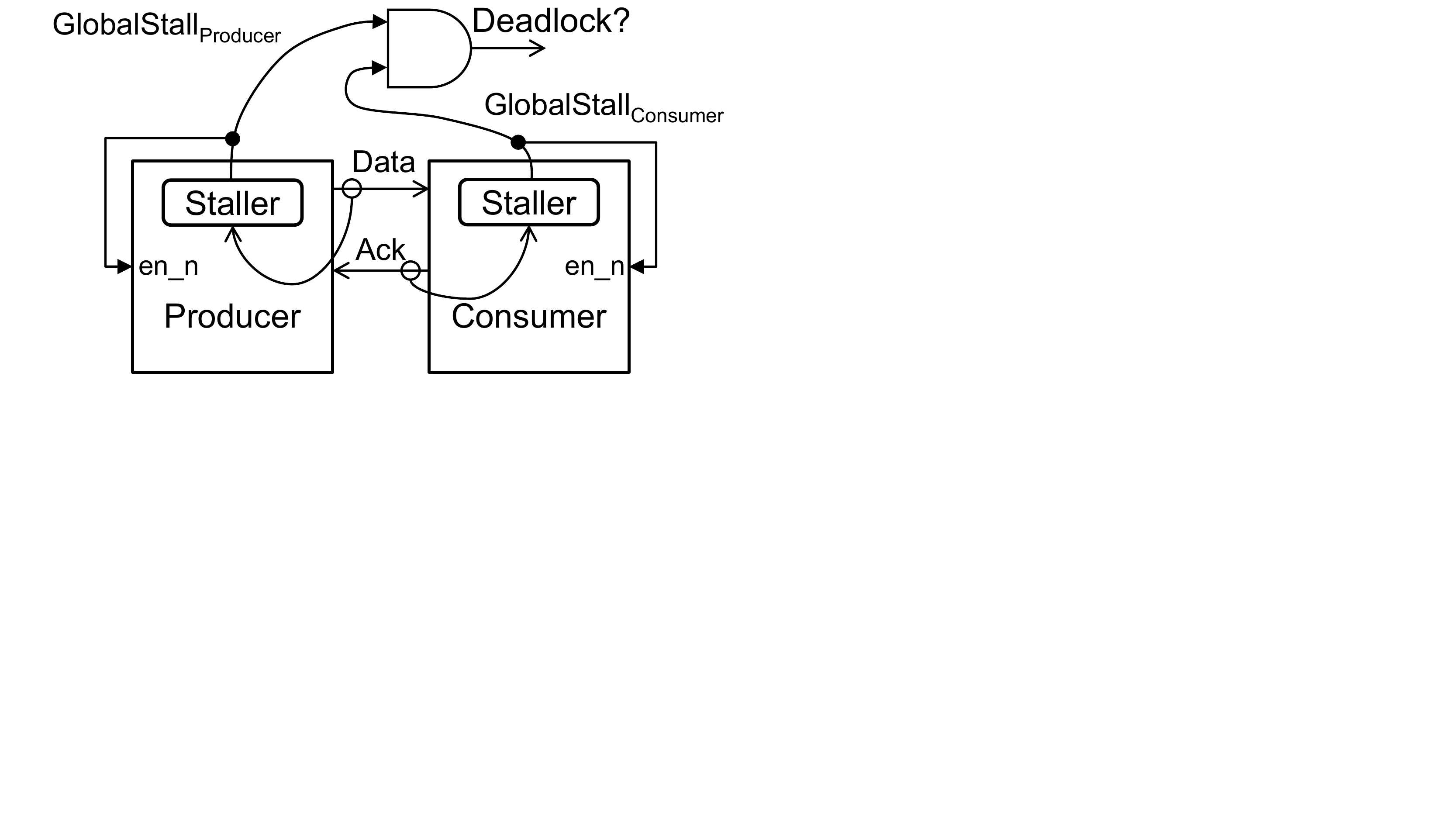}
\vspace{0.05in}
\caption{\BF{Formal model for verifying deadlock-free property} --- Verifies whether all modules are in the stalled state simultaneously.}
\label{fig:deadlock_check}
\end{figure}

Our deadlock detection proves whether a design is absent of deadlock.
Figure~\ref{code:simple_foobar} shows a simple system with two interacting functional blocks that contains a potential deadlock.
If \texttt{cond} in the \texttt{Consumer} function initializes to false, consumer never accepts the data from producer, but instead immediately tries to push an acknowledgment to producer.
However, producer would stall during push because its data are not accepted and cannot move on to popping the acknowledgment from consumer.
As a result, both producer and consumer end up in a stalled state.
When all interacting modules are stalled, no module can trigger forward progress, and the system is in deadlock.
This represents one of many scenarios for which an LI design could deadlock.

\begin{figure}[b]
  \vspace{-3pt}
  \hspace{18pt}
  \begin{minipage}[c]{0.4\columnwidth}
    \centering
    \small
\begin{lstlisting}[language=C++,columns=fullflexible]
void Producer() {
  while (1) {
   Data.Push(1);      
   Ack.Pop();
}}
\end{lstlisting}
  \end{minipage}%
  \begin{minipage}[c]{0.5\columnwidth}
    \centering
    \small
\begin{lstlisting}[language=C++,columns=fullflexible]
void Consumer() {
  while (1) {
   if (cond) {Data.Pop();}      
   Ack.Push(1);
}}
\end{lstlisting}
  \end{minipage}
\caption{\BF{Simple system in C++ with two interacting functional blocks} --- Both blocks will stall if \TT{Consumer} does not accept data from \TT{Producer}.}
\label{code:simple_foobar}
\end{figure}

To detect a deadlock under any of these scenarios, we can make use of global stall signals that a typical HLS tool generates for individual RTL modules.
For instance, Mentor Catapult HLS assigns a wait controller for each stallable interface (e.g., blocking port) of a module. 
These wait controllers then communicate with the staller of the module to form the global stall signal for clock gating the module~\cite{graphics2020catapult}.
Figure~\ref{fig:deadlock_check} provides an abstract illustration of the producer and consumer modules in RTL generated from the C++ descriptions in Figure~\ref{code:simple_foobar}.
As shown in Figure~\ref{fig:deadlock_check}, our verification wrapper constructs the formal model by aggregating the global stall signals from individual modules.
During the formal check, our model ensures that any input or output to the system is not blocked.
The system contains a deadlock if the formal check determines that it is possible for all aggregated global stall signals to be asserted at the same time.
If not, the system is free of deadlock.
The deadlock-free property for an $N$-module design is formally expressed as $\neg\left(\bigwedge_{i=1}^{N} GlobalStall_i\right)$.

We extend our deadlock detection method to support non-blocking ports, which do not include explicit stall signals because they cannot be blocked by definition.
In this case, we devise a custom global stall signal for each RTL module from the ready signals of its non-blocking ports.
This custom global stall signal is asserted if none of the relevant ready signals have been asserted for $N$ clock cycles, where $N$ is a known constant at verification time based on the HLS-applied optimizations.
This type of custom global stall signal can be used in lieu of the global stall signal in Figure~\ref{fig:deadlock_check} to implement the same deadlock check.
\section{Experiments}
\label{sec-experiments}

Our formal models are implemented using SystemVerilog assertions and verified with bounded model checking~\cite{biere2003bounded} using Synopsys VC Formal 2018.10 running on an Intel Xeon CPU at 3GHz.
Formal model checking is performed on the RTL synthesized from C++ using HLS tool.
Because HLS is an automated process that compiles C++ into predictably-structured RTL well-suited for the extraction of relevant signals outlined in Sections~\ref{subsec-invalid} and~\ref{subsec-deadlock}, our formal flow can be fully automated.
Although the bugs we target originate in C++ during design entry, we formally verify the designs in RTL because the full scopes of the bugs only manifest under the  cycle-accurate timing model of RTL.

\begin{table}[t]
\footnotesize
\setlength{\tabcolsep}{0.5em}
\caption{\BF{Results of formal modle checking on realistic abstracted bug cases} --- Runtimes are reported in seconds.}
\begin{center}
\begin{tabular}{|l|l|l|l|l|l|}
\hline
\multicolumn{1}{|c|}{\textbf{Design}} & \multicolumn{1}{c|}{\textbf{Model}} & \multicolumn{1}{c|}{\textbf{Version}} & \multicolumn{1}{c|}{\textbf{\#Gates}} & \multicolumn{1}{c|}{\textbf{Runtime}} & \multicolumn{1}{c|}{\textbf{Result}} \\ \hline
\multirow{2}{*}{\shortstack[l]{Unconstrained \\Input}} & \multirow{2}{*}{\shortstack[l]{Invalid \\Input}} & Initial & 0.1k & 5.99 & Falsified \\ \cline{3-6} 
 &  & 1st fix & 0.1k & 6.94 & Proven \\ \hline 
\multirow{3}{*}{\shortstack[l]{Under-constrained \\Read}} & \multirow{3}{*}{\shortstack[l]{Invalid \\Input}} & Initial & 1.9k & 14.14 & Falsified \\ \cline{3-6} 
 &  & 1st fix & 1.9k & 18.24 & Falsified \\ \cline{3-6} 
 &  & 2nd fix & 1.9k & 359.3 & Proven \\ \hline
\multirow{3}{*}{\shortstack[l]{Out-of-order Push}} & \multirow{3}{*}{\shortstack[l]{Deadlock}} & Initial & 0.8k & 12.87 & Falsified \\ \cline{3-6} 
 &  & 1st fix & 0.8k & 12.71 & Falsified \\ \cline{3-6} 
 &  & 2nd fix & 0.8k & 32.08 & Proven \\ \hline
\multirow{2}{*}{\shortstack[l]{Circular \\Dependency}} & \multirow{2}{*}{\shortstack[l]{Deadlock}} & Initial & 1.2k & 13.51 & Falsified \\ \cline{3-6} 
 &  & 1st fix & 1.2k & 17.68 & Proven \\ \hline
\multirow{2}{*}{\shortstack[l]{Mismatched \\Pipeline Depths}} &
\multirow{2}{*}{\shortstack[l]{Deadlock}} & Initial & 0.7k & 8.24 & Falsified \\ \cline{3-6} %
 &  & 1st fix & 0.7k & 18.47 & Proven \\ \hline 
\end{tabular}
\label{tbl:val_results}
\end{center}
\end{table}

We first validate our formal models using known bug cases, listed in Table~\ref{tbl:val_results}, abstracted from real industry and academic HLS designs.
We abstract the design names to indicate the primary cause of the bugs.
Each of these bug cases consists of the initial (buggy) version of the corresponding design followed by one or more patched (but possibly still buggy) versions of the same design.
The initial versions of all the designs are written without knowledge of our formal models.
Likewise, the models are developed without specific knowledge of the initial designs.
As such, our abstracted bug cases provide a minimally viable but faithful reproduction of the specific bugs to help narrow down the root causes of the bugs and to understand and validate the results of our formal models.

Table~\ref{tbl:val_results} details the findings of the formal engine against our proposed models for different versions of each design, along with post-logic-synthesis gate count and runtime of the formal engine.
The evolution of each design from the initial version to the final fix demonstrates the effectiveness and correctness of our formal models.
As shown in the table, our formal models have extracted bugs even from purportedly fixed and verified designs, which speaks to the shortcoming of the existing verification methodology.
The counterexamples provided by these proofs are instrumental in quickly identifying the root cause and devising the appropriate fix.

To further demonstrate the applicability of our approach, we apply our formal models on a set of open-source HLS library components from MatchLib~\cite{khailany2018modular}.
These library components are meant to be reused across a large number of designs, and therefore constitute good candidates for extensive verification.
We also experiment with full applications of our own and from open-source HLS benchmark suite Rosetta~\cite{zhou2018rosetta} to demonstrate the general applicability of our models.
As shown in Table~\ref{tbl:additional_results}, we apply the invalid input check on the five design components and successfully prove that they are unaffected by invalid inputs.
On the other hand, we apply deadlock detection on a 2x2 array of network-on-chip (NoC) router components, an optical flow accelerator, and a machine-learning accelerator for spam filtering, where we identify certain deadlock states.

Results in Table~\ref{tbl:additional_results} shows that we are able to prove exhaustively that the five components are unaffected by invalid input, and prove at the user-defined bound that specific versions of the applications are free of deadlock.
Specifically, our deadlock check supports the NoCRouterArray application which makes use of non-blocking ports exclusively. We discover an incorrect protocol constraint in NoCRouterArray.v1 that results in a deadlock and apply the appropriate fix in NoCRouterArray.v2 after examining the counterexample trace. 
We also detect a deadlock in OpticalFlow.v2 that escapes the test cases in the supplied test bench.
Compared to OpticalFlow.v1, OpticalFlow.v2 is buggy because it contains one FIFO with a reduced capacity, resulting in a potential cyclic wait scenario from insufficient FIFO capacity due to the re-convergence pattern in the benchmark's dataflow~\cite{fingeroff2010high}.
SpamFilter.v2 does not incur the same problem even with reduced FIFO sizes compared to SpamFilter.v1 because the benchmark's dataflow pattern contains no branches and thus no re-convergence.

\begin{table}[t]
\footnotesize
\setlength{\tabcolsep}{0.35em}
\caption{\BF{Results of formal model checking on HLS benchmarks} --- Runtimes are reported in seconds. Bounds are shown under Results/Bound for non-exhaustive proofs.}
\begin{minipage}[c]{\columnwidth}
\begin{center}
\begin{tabular}{|c|l|l|l|l|l|}
\hline
& \multicolumn{1}{c|}{\textbf{Design}} & \multicolumn{1}{c|}{\textbf{Model}} &
\multicolumn{1}{c|}{\textbf{\#Gates}} & \multicolumn{1}{c|}{\textbf{Runtime}} & \multicolumn{1}{c|}{\textbf{Result/Bound}} \\ \hline
\multirow{5}{*}{\rotatebox[origin=c]{90}{Components}} & 5-input Arbiter & Invalid input & 0.10k & 10.98 & Proven \\ \cline{2-6} 
& 16-bit Adder4 & Invalid input & 0.77k & 30389 & Proven \\ \cline{2-6} 
& AxiMasterGate  & Invalid input & 6.63k & 6328 & Proven \\ \cline{2-6} 
& AxiSlaveToReg  & Invalid input & 58.7k & 36627 & Proven \\ \cline{2-6} 
& WHVCRouter & Invalid input & 9.75k & 18250 & Proven \\ \hline 
\multirow{6}{*}{\rotatebox[origin=c]{90}{Applications}} & NoCRouterArray.v1 & Deadlock & 32.0k & 12.71 & Falsified \\ \cline{2-6} 
& NoCRouterArray.v2 & Deadlock & 32.0k & 16915 & 450 \\ \cline{2-6} 
& OpticalFlow.v1\textsuperscript{$\star$} & Deadlock & 321k\textsuperscript{+} & 42090 & 10000 \\ \cline{2-6} 
& OpticalFlow.v2\textsuperscript{$\dagger$} & Deadlock & 321k\textsuperscript{+} & 1237 & Falsified \\ \cline{2-6} 
& SpamFilter.v1\textsuperscript{$\star$} & Deadlock & 140k\textsuperscript{+} & 15341 & 10000 \\ \cline{2-6}  
& SpamFilter.v2\textsuperscript{$\diamond$} & Deadlock & 140k\textsuperscript{+} & 10548 & 10000 \\ \hline 

\end{tabular}
\end{center}
\label{tbl:additional_results}
\end{minipage}
\begin{center}
\begin{minipage}[l]{0.9\columnwidth}
\vspace{-10pt}
\textsuperscript{$\star$}Original benchmark.  
\textsuperscript{$\dagger$}All HLS FIFO depths=1024.
\textsuperscript{$\diamond$}Maximum HLS FIFO depth=4.
\textsuperscript{+}Gate count excludes 9kB-269kB of SRAMs.
\end{minipage}
\end{center}
\end{table}

\section{Conclusions}
\label{sec-conclusions}

While the LI design methodology simplifies the composition of synthesized functional blocks in HLS, this latency-tolerant design style also introduces additional difficulty in ensuring the correctness of high-level designs.
In this paper, we provide an automated flow based on formal model checking to guarantee that a high-level LI design is not affected by invalid input data and is free of deadlock.
The proposed verification techniques are effective and generally applicable to a range of LI-based HLS designs, and lead to promising improvement in the quality of verification.
We believe that closing the verification gap is key to mainstream adoption of HLS tools, and our formal verification extensions play a crucial role as part of static sign-off toward this direction~\cite{ashar2019closing}.
We expect wider adoption of our approach as formal tools and their underlying engines become increasingly scalable~\cite{mann2020partial,barrett2011cvc4}.

\bibliographystyle{IEEEtran}
\bibliography{IEEEabrv,refs.bib}

\end{document}